\documentclass[letterpaper]{sae}
\usepackage{textcomp}
\usepackage{cite}
\usepackage{multirow}
\usepackage{amsmath}
\usepackage{float}
\usepackage{booktabs}
\usepackage{siunitx}
\usepackage{tikz}
\usepackage{hyperref}
\hypersetup{colorlinks=true,linkcolor=black,citecolor=black,urlcolor=black}

\usepackage{caption}            
\usepackage[table]{xcolor}
\usepackage{graphicx} 

\PaperTitle{Safety Monitor for Off-Road Planning with Uncertainty Bounded Bekker Costs}
\AddAuthor{Akshay Naik$^{*}$, Ramavarapu S. Sreenivas, William R. Norris$^{\dagger}$}{University of Illinois Urbana-Champaign, Urbana, IL, USA}
\AddAuthor{Albert E. Patterson}{Texas A\&M University, College Station, TX, USA}
\AddAuthor{Ahmet Soylemezoglu, Dustin Nottage}{U.S. Army Corps of Engineers Engineering Research and Development Center, Champaign, IL, USA}
\PaperNumber{26IDM-0025}
\SAECopyright{2026}

\usepackage{hyperref}
\hypersetup{
    colorlinks=true,
    linkcolor=blue,
    citecolor=blue,
    filecolor=magenta,      
    urlcolor=cyan,
}

\begin{document}
\maketitle

\begingroup
\renewcommand{\thefootnote}{\fnsymbol{footnote}} 
\footnotetext[1]{Corresponding author - Email: akshayn3@illinois.edu}
\footnotetext[2]{Deceased - contributed significantly to the initial phases of this work.}
\endgroup

\section{Abstract}
Reliable off-road autonomy requires operational constraints so that behavior stays predictable and safe when soil strength is uncertain. This paper presents a runtime assurance safety monitor that collaborates with any planner and uses a Bekker-based cost model with bounded uncertainty. The monitor builds an upper confidence traversal cost from a lightweight pressure sinkage model identified in field tests and checks each planned motion against two limits: maximum sinkage and rollover margin. If the risk of crossing either limit is too high, the monitor switches to a certified fallback that reduces vehicle speed, increases standoff from soft ground, or stops on firmer soil. This separation lets the planner focus on efficiency while the monitor keeps the vehicle within clear safety limits on board. Wheel geometry, wheel load estimate, and a soil raster serve as inputs, which tie safety directly to vehicle design and let the monitor set clear limits on speed, curvature, and stopping at run time. The method carries uncertainty analytically into the upper confidence cost and applies simple intervention rules. Tuning of the sinkage limit, rollover margin, and risk window trades efficiency for caution while keeping the monitor light enough for embedded processors. Results from a simulation environment spanning loam to sand include intervention rates, violation probability, and path efficiency relative to the nominal plan, and a benchtop static loading check provides initial empirical validation.

\section{Introduction}
Off-road vehicles operate in environments that rarely follow neat or repeatable patterns. Soil strength, moisture, and texture can vary within a few meters, and slopes often interact with traction in ways that are difficult to predict. These unstructured conditions appear across multiple domains, including agriculture, construction, defense, and planetary exploration. In each case, reliable mobility depends on anticipating how the terrain will respond before the vehicle commits to a path \cite{fankhauser_probabilistic_2018, leung_infusing_2020, underhill_ensuring_2025}. Modern path planners have improved efficiency in rough terrain, but most rely on deterministic cost maps or learning-based estimates that lack formal uncertainty bounds \cite{jerome_real-time_2025, bhuiyan_deep-reinforcement-learning-based_2023}. When a planner overestimates traction or underestimates sinkage, the vehicle can exceed safe limits, resulting in loss of control or immobilization. For on-road vehicles, this uncertainty is small and predictable, while for off-road vehicles it is structural and unavoidable. The primary challenge is that the soil response depends on variables that cannot be measured directly in real time. Classical terramechanics models, such as Bekker’s pressure-sink relationship and Wong’s stress-shear formulations \cite{bekker_off--road_1960, wong_prediction_1967}, describe the interaction of the terrain with clear physical interpretability. These models relate load, contact area, and soil deformation through empirically derived constants \cite{bekker_introduction_1969, wong_prediction_1967, wong_theory_2001}. However, their use in autonomous systems has been limited because the relevant field parameters are uncertain and often vary with moisture, compaction, or wheel slip. Without explicitly incorporating this uncertainty into planning, even well-calibrated models can produce unreliable predictions in unstructured terrain \cite{jia_terramechanics-based_2012}.  

To address this limitation, this paper proposes a safety monitor that applies a Bekker-based cost model with bounded uncertainty to define reliable operational limits. The natural variability in soil properties is represented as a bound $\pm$20\% around the nominal parameter values, consistent with the experimental observations from previous terramechanics studies \cite{shamrao_estimation_2018}. The monitor generates an upper-confidence estimate of the traversal cost from these bounded parameters and checks each planned motion against sinkage and rollover margins. If the probability of exceeding either limit becomes too high, the monitor intervenes by slowing the vehicle, increasing standoff distance from soft ground, or stopping movement on firmer terrain. This approach is based on reliability-based design and runtime assurance principles \cite{phan_component-based_2017, trevisan_dynamic_nodate, knoedler_safety_2025}, allowing the planner to focus on efficiency while the monitor enforces safety constraints directly linked to terrain physics. The following sections describe the soil-vehicle model used to form uncertainty-bounded costs, the structure of the runtime safety monitor, and its evaluation within a simulation environment spanning soft to firm soils. The results illustrate how varying uncertainty bounds affect intervention rates, violation probability, and efficiency tradeoffs, followed by an initial empirical validation through static loading tests.  

\section{Background}
Safety assurance in autonomous systems depends on runtime checks and fallback routines that keep operation within safe limits as uncertainty grows. Early examples, such as the Simplex architecture, showed that an autonomous controller could operate efficiently under normal conditions, but hand control back to a verified safety channel when confidence fell \cite{phan_component-based_2017}. These principles later influenced how robotics and vehicle systems separate performance logic from safety layers. Until recently, most safety frameworks focused narrowly on controlling or sensing faults, rarely considering how terrain itself can introduce uncertainty. Functional safety standards such as ISO 26262 and SOTIF have begun to address this broader view by encouraging explicit treatment of uncertainty during decision-making \cite{leung_infusing_2020, underhill_ensuring_2025}.

Terrain modeling has followed a parallel evolution driven by the need to predict traction, slippage, and sinkage before motion. The early approaches were built on the Bekker pressure-sinkage model and the shear deformation principles formalized by Wong \cite{bekker_introduction_1969, wong_theory_2001}. These models remain central to terramechanics because they connect measurable quantities such as wheel load and contact area to soil deformation through interpretable parameters. Their weakness lies in parameter sensitivity; soil constants vary sharply with moisture, compaction, and grain composition, often changing faster than field estimation can. Probabilistic terrain mapping sought to address this limitation by learning local traversability distributions from sensor data \cite{fankhauser_probabilistic_2018}. More recent extensions incorporated uncertainty propagation and Gaussian process modeling to quantify confidence in each predicted cost layer \cite{torroba_fully-probabilistic_2022}. Although these data-driven methods improve adaptability, they typically treat uncertainty statistically, not analytically, and do not enforce physical constraints derived from the soil–vehicle interaction itself.

Bridging safety assurance and terrain modeling remains an emerging area. Trevisan et al. introduced a dynamic reachability assurance framework that applies risk-aware constraints during planning \cite{trevisan_dynamic_nodate}. Knoedler et al. proposed the construction of safety filters at runtime using policy control barrier functions to maintain stability and constraint satisfaction under uncertainty \cite{knoedler_safety_2025}. These approaches move toward onboard reliability management but rely on abstract risk metrics that do not explicitly connect to terrain physics. They safeguard motion feasibility, but not ground interaction limits such as sinkage or rollover. This gap motivates a combined perspective. Although prior work on runtime assurance formalized safety logic for decision-making and terrain research improved uncertainty estimation, few efforts have merged the two physically grounded. The safety monitor developed in this article builds directly on the soil mechanics of Bekker and Wong while analytically incorporating uncertainty into the cost model through bounded physical margins. By bounding the traversal cost through upper-confidence estimates of sinkage and traction, the monitor provides a physically interpretable reliability layer that complements existing planners rather than replacing them. This approach allows safety to be enforced not through abstract risk thresholds, but through measurable terrain response linked to vehicle geometry and loading.

\section{Methodology}
This section describes how uncertainty in soil behavior is modeled, evaluated, and enforced at run time through a lightweight safety monitor. The approach starts by expressing the traversal cost using the Bekker pressure-sinkage model and bounding that cost according to the uncertainty in the soil parameters \cite{bekker_introduction_1969, wong_theory_2001}. The resulting cost map, combined with slope information from the elevation grid, allows each planned motion to be evaluated against physical limits on sinkage and rollover. The runtime safety monitor evaluates each path generated by the motion planner and intervenes when either limit is exceeded. This framework is designed to operate alongside any planner that outputs discrete waypoints on a cost map. In this implementation, it is integrated with a Vehicle-Dynamic RRT* planner \cite{naik_hybrid_2025} for demonstration.

\begin{table*}[h!]
    \centering
    \small
        \renewcommand{\arraystretch}{1.15}
    \caption{Nominal Bekker parameters for representative soil types.}
    \label{tab:soil_params}
    \begin{tabular}{>{\centering\arraybackslash}p{0.227\linewidth} >{\centering\arraybackslash}p{0.227\linewidth} >{\centering\arraybackslash}p{0.227\linewidth} >{\centering\arraybackslash}p{0.227\linewidth}}
    \hline \rowcolor{lightgray}
    \textbf{Soil Type} & $k_c$ [N\,m$^{-(n+1)}$] & $k_\phi$ [N\,m$^{-(n+2)}$] & $n$ \\
    \hline
    Pavement (compacted) & $1.0\times10^6$ & $1.0\times10^7$ & 1.0 \\
    Gravel (3–6\,mm) & 0 & $5.0\times10^5$ & 1.0 \\
    Wood chips (dry) & $7.0\times10^3$ & $1.5\times10^6$ & 0.8 \\
    Loam / field dirt & $1.0\times10^3$ & $1.8\times10^6$ & 1.0 \\
    Grass (moist) & $1.0\times10^3$ & $1.2\times10^6$ & 0.9 \\
    Loose dune sand & $2.0\times10^3$ & $5.0\times10^5$ & 1.2 \\
    \bottomrule
    \end{tabular}
\end{table*}

\textbf{Modeling Bounded Bekker Costs}: The terrain cost model is based on Bekker’s pressure–sinkage relationship, which relates normal pressure \(p\) to soil deformation \(z\) as
    \begin{equation}
        p \;=\; \Big(\frac{k_c}{b} + k_\phi\Big)\, z^{\,n},
        \label{eq:bekker}
    \end{equation}
where \(b\) is the contact width and \(k_c, k_\phi, n\) are empirical soil parameters. Representative values for several soil types used in the simulation are shown in Table~\ref{tab:soil_params} \cite{naik_hybrid_2025}. Each cell in the terrain grid stores a discrete soil label that maps to these nominal parameters. For a wheel of radius \(R\) and normal load \(W\), the average contact pressure is:
    \begin{equation}
        \begin{aligned}
            a &= \sqrt{2Rz - z^2}, &\qquad \ell &= 2a,\\
            A &\approx b\,\ell = 2b\,a, &\qquad p_\text{avg} &= \frac{W}{A}.
        \end{aligned}
        \label{eq:contact_pressure}
    \end{equation}
Equating \(p_\text{avg}\) with \ref{eq:bekker} yields the sinkage $z$ for a given cell. The traversal cost for that cell is then defined from the predicted deformation; in this work, it is proportional to the conservative (upper-confidence) sinkage described below. Because soil parameters vary with composition, moisture, and compaction, they are modeled with bounded uncertainty. Each parameter is assigned a uniform margin \(\delta\) about its nominal value:
    \[
    k_c^\pm = k_c(1 \pm \delta), \quad
    k_\phi^\pm = k_\phi(1 \pm \delta), \quad
    n^\pm = n(1 \pm \delta),
    \]
with \(\delta=0.20\) in this implementation. The monitor forms a conservative sinkage by maximizing the predicted \(z\) over this bounded set,
    \begin{equation}
        z_{uc} \;=\; \max_{k_c,k_\phi,n}\,
        \left(\frac{p_\text{avg}}{k_c/b + k_\phi}\right)^{\!1/n},
        \label{eq:uncertainty}
    \end{equation}
The upper-bounded traversal cost \(C_{uc}\) is then defined as a monotone function of \(z_{uc}\):
    \begin{equation}
        C_{uc} \;=\; f(z_{uc}).
        \label{eq:uncertainty_cost}
    \end{equation}
This construction preserves an analytic link between parameter uncertainty and risk while remaining light enough for embedded execution. In addition to the sinkage-based cost, the model includes a rollover term derived from the local slope. The slope is estimated from the elevation gradient \(s=\|\nabla h\|\) and optionally mapped to a roll angle \(\theta=\arctan(s)\). This deterministic term penalizes steep regions where lateral stability decreases and is evaluated alongside \(C_{uc}\) by the safety monitor.
    
\begin{figure*}[h!]
    \centering
    \includegraphics[width=0.8\linewidth]{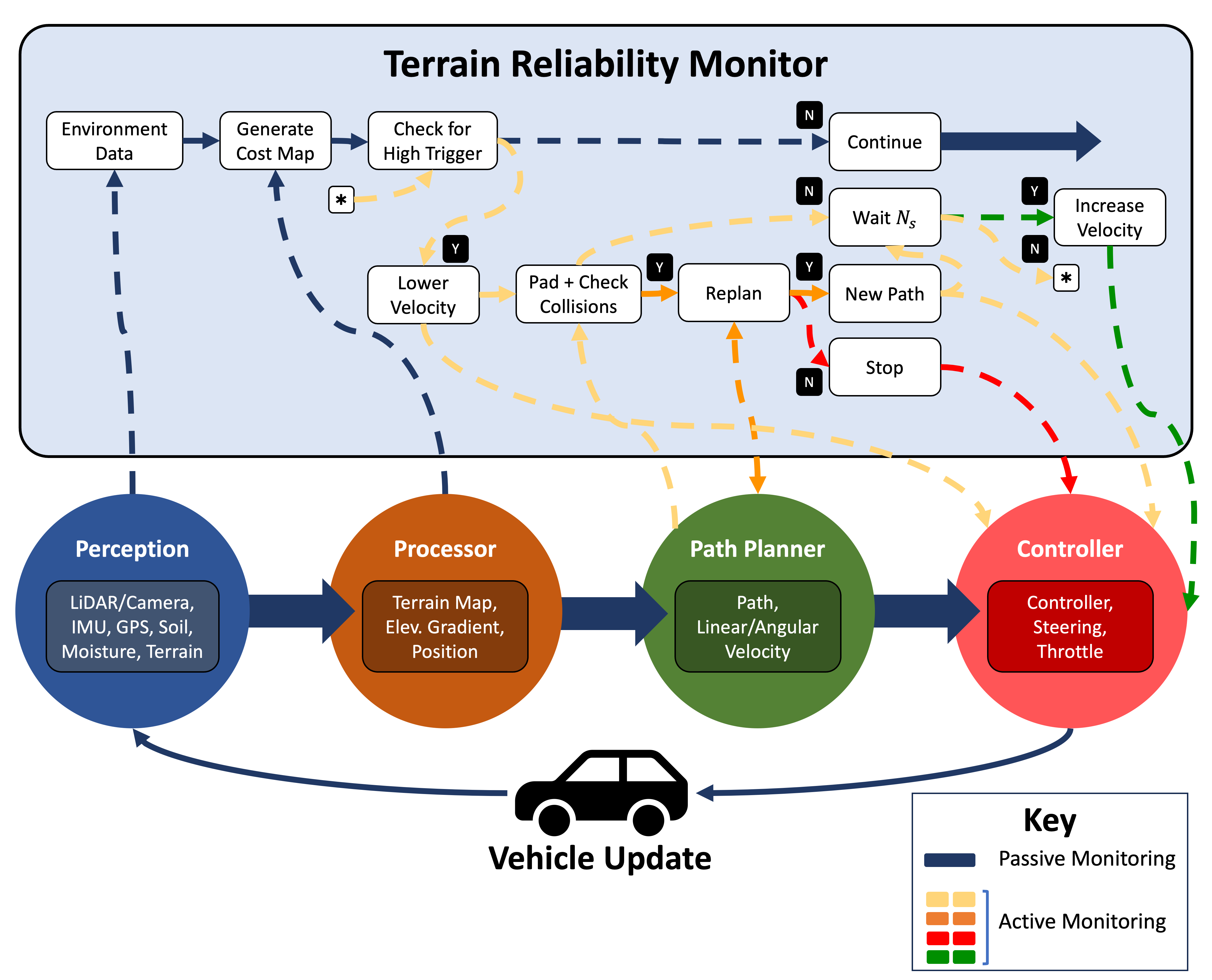}
    \caption{System overview showing integration of the Terrain Reliability Monitor (TRM) within the autonomous ground vehicle pipeline. The TRM observes perception, mapping, and planning outputs to detect unsafe terrain, apply $v_{min}$ velocity limits, pad high-cost regions, and trigger re-plans when necessary. Once stability is maintained for $N_s$ updates, the TRM returns to passive monitoring.}
    \label{fig:monitor_framework}
\end{figure*}

\textbf{Runtime Safety Monitor}: The runtime safety monitor evaluates planned motion in real time to ensure that vehicle behavior remains within the prescribed limits on sinkage and rollover~\ref{fig:monitor_framework}. It operates as a supervisory layer running in parallel with the motion planner, intercepting the current path, the local terrain map, and the vehicle state in each planning cycle. The primary inputs of the monitor are the upper-bounded sinkage cost \(C_{uc}\) derived from Equations \ref{eq:bekker}–\ref{eq:uncertainty_cost}, the slope-based rollover cost \(C_{roll}\) from (6) and the geometric and loading parameters of the vehicle. Terrain costs are defined over a discrete grid, and the active path is checked cell by cell along its sampled waypoints. In each cycle, the monitor scans the vehicle’s sensing window for high-cost terrain. high-cost detection occurs if any detected cell satisfies:
    \[
    C_{uc} > C_{sink,\max} \quad \text{or} \quad C_{roll} > C_{roll,\max}.
    \]
The thresholds \(C_{sink,\max}\) and \(C_{roll,\max}\) correspond to the maximum allowed sinkage \(z_{max}\) and the critical rollover slope \(\theta_{max}\) determined from the geometry of the wheel and the height of the vehicle's center of gravity. 

Suppose that high-cost terrain is detected anywhere within the sensing window. In that case, the vehicle immediately reduces the speed to the minimum safe value \(v_{min}\), regardless of whether the current path intersects the region. This reduction maximizes safety by limiting sinkage growth and providing additional time for terrain updates. The vehicle resumes its nominal speed \(v_{nom}\) only after \(N_s\) consecutive updates in which no high-cost cells are detected. Upon detection, the monitor expands the high-cost region by a kinematic buffer distance \(d_{pad}\) to ensure sufficient clearance for an evasive turn. The padding is derived from the vehicle’s minimum turning radius and footprint:
    \begin{equation}
        \begin{aligned}
        R_{\min} &= \frac{L}{\tan\delta_{\max}}, \\
        d_{pad} &= R_{\min} + \frac{w}{2} + b, \\
        n_{pad} &= \left\lceil \frac{d_{pad}}{r} \right\rceil .
        \label{eq:padding}
        \end{aligned}
    \end{equation}
where \(L\) is the wheelbase, \(\delta_{\max}\) is the maximum steering angle, \(w\) is the vehicle width, \(b\) is a small buffer for grid discretization, and \(r\) is the grid cell size. If the sensing window cannot accommodate the required padding (\(d_{pad}\)), the vehicle halts immediately. If the current path intersects the padded region, the monitor triggers a re-plan while the vehicle continues at \(v_{min}\). The available time before reaching the padded boundary is:
    
    \begin{equation}
        \begin{aligned}
        d_{\downarrow} &= \frac{(v_k - v_{min})^2}{2\,a_{\text{brake}}}, \\
        t_{safe} &= \max\!\Big(0,\frac{d_{margin} - d_{\downarrow}}{v_{min}}\Big)
                   - \tau_{\text{comp}},
        \label{eq:t_safe}
        \end{aligned}
    \end{equation}
    
where \(v_k\) is the current speed, \(a_{\text{brake}}\) is the deceleration limit, \(\tau_{\text{comp}}\) accounts for computation and actuation latency, and \(d_{margin}\) is the distance to the padded boundary along the current heading. If a new path is received before \(t_{safe}\) expires, the vehicle follows it at \(v_{min}\). If \(t_{safe} \le 0\) or the re-plan fails within this time window, the vehicle halts at the last safe cell. From this state, external actions, such as switching to another planner, enabling reverse motion, or manual control, may occur while the monitor continues to enforce the same safety logic. This layered sequence (slowdown on detection, buffer expansion, time-bounded re-plan, and halt) ensures containment of risk even under bounded uncertainty in terrain parameters. All decisions are computed locally, enabling real-time operation on embedded hardware.

\begin{figure*}[h!]
    \centering
    \includegraphics[width=0.77\linewidth]{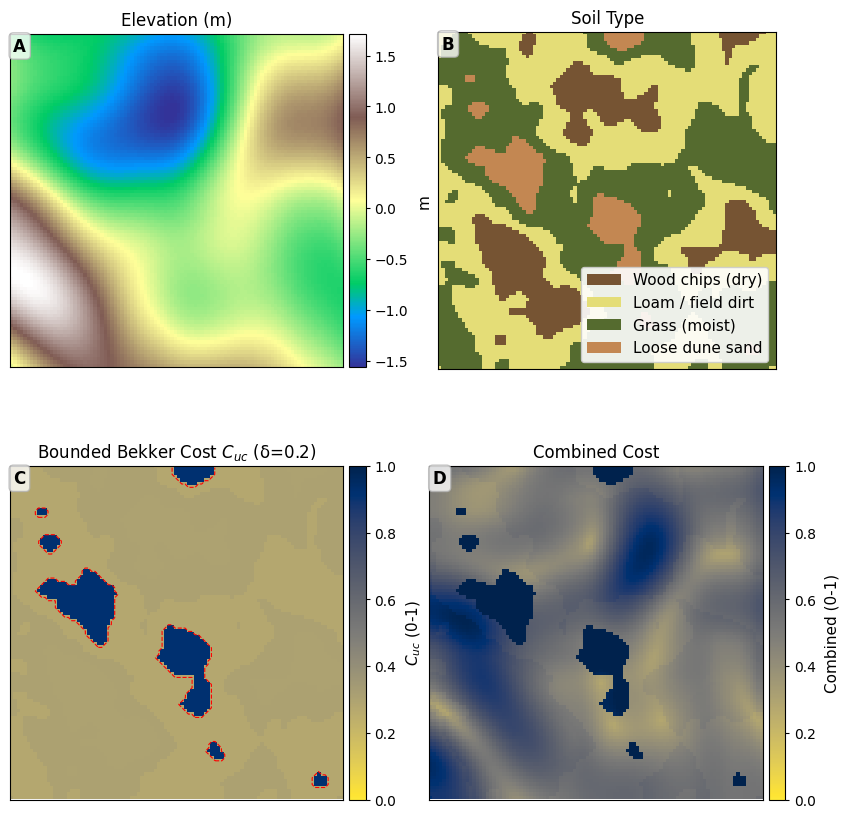}
    \caption{Terrain and cost maps used in simulation. 
    (A) Elevation used to compute $C_{roll}$ from slope. 
    (B) Soil-type map assigning integer labels for Bekker parameter lookup. 
    (C) Bounded Bekker cost $C_{uc}$ computed with $\delta = 0.2$. 
    (D) Combined terrain cost $C_{total} = C_{uc} + \lambda C_{roll}$ with $\lambda = 2.5$.}
    \label{fig:terrain_cost_maps}
\end{figure*}

\textbf{Integration with Planner and Simulation Setup}: The safety monitor is designed to operate alongside any motion planner that outputs a discrete sequence of waypoints over a cost map. For evaluation, it is integrated with a Vehicle-Dynamic RRT* (VD-RRT*) planner \cite{naik_hybrid_2025} that models the vehicle using a bicycle kinematic structure and computes tire forces using Pacejka’s Magic Formula. This configuration ensures consistency between the terrain-dependent cost model and the dynamic feasibility of the vehicle. At each iteration, the planner proposes a dynamically feasible path from the current pose of the vehicle to the goal while minimizing the total cost of the traverse. The edge cost of the planner is defined as the cumulative sum of \(C_{uc} + \lambda C_{roll}\) along each sampled trajectory, where \(\lambda\) weights the slope sensitivity relative to the risk of sinking. Control inputs are bounded by the steering-rate and acceleration limits consistent with the vehicle’s dynamic model, to ensure that the motions sampled remain physically realizable. The monitor evaluates each candidate and active path using the bounded Bekker cost $C_{uc}$ and the slope-based rollover cost $C_{roll}$, both of which are supplied to the planner in real time. Instead of duplicating the internal logic described in Section~III-B, this subsection emphasizes the data flow between components. The planner proposes a feasible path; the monitor then checks it against terrain costs, adjusting velocity, or initiating a re-plan as needed. Figure~\ref{fig:monitor_framework} summarizes this exchange, highlighting the continuous loop of sensing, evaluation, and intervention that governs safe navigation.

\begin{figure*}[h!]
    \centering
    \includegraphics[width=0.75\linewidth]{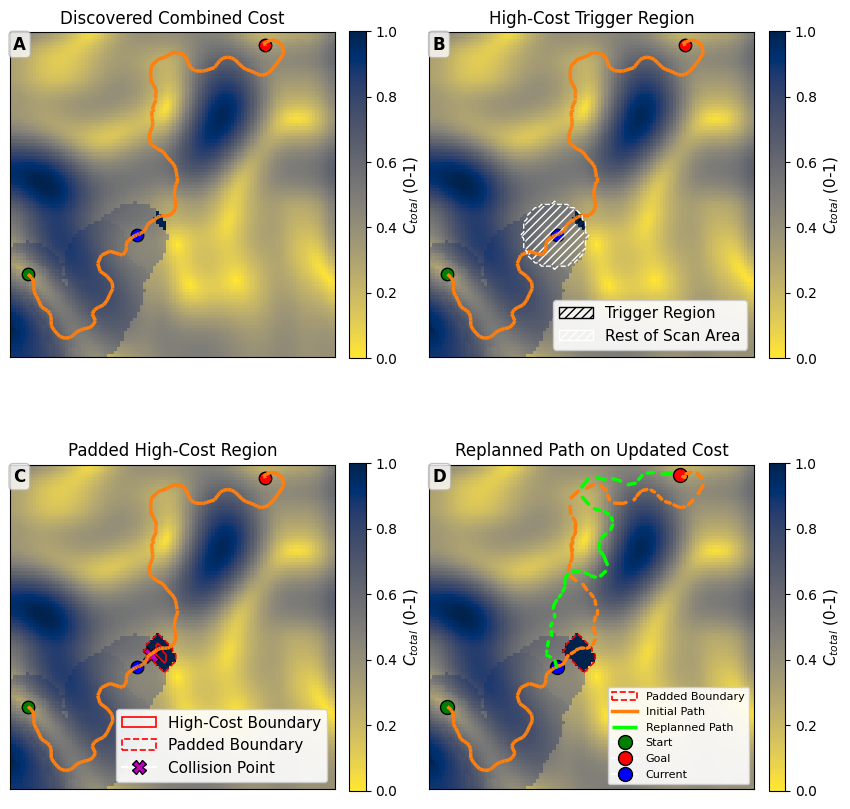}
    \caption{Runtime monitor intervention sequence. 
    (A) True cost map revealed through sensing. 
    (B) Detection of high-cost region. 
    (C) Padded boundary based on vehicle escape distance. 
    (D) Re-plan around unsafe region before timeout.}
    \label{fig:trigger_cost_maps}
\end{figure*}

The vehicle operates within a discretized map where each grid cell stores the soil parameters and elevation data required to compute \(C_{uc}\) and \(C_{roll}\). Terrain variability is introduced through Perlin noise to emulate spatially correlated transitions between soft and firm soils. The elevation map is smoothed with a \(3\times3\) kernel to approximate gradual slope variations rather than abrupt discontinuities. During simulation, the vehicle starts from an initial path computed with the nominal cost map, then progressively updates its perception as new terrain is sensed within a fixed radius around its current position. To model sensing constraints, the vehicle’s perception range is limited to a local window centered on its pose, representing onboard LiDAR or stereo depth coverage. Each update reveals the true terrain conditions within that window, prompting a local recalculation of cost and a possible re-plan according to the monitor’s logic. This closed-loop cycle continues until the goal is reached or the vehicle halts because of excessive risk.The monitor and planner are implemented in Python for analysis and all computations run at 10~Hz in simulation. Although the presented implementation uses VD-RRT* \cite{naik_hybrid_2025}, the same supervisory structure can operate with other planners such as D*, A*, or lattice-based methods, provided that they output waypoints compatible with the discretized terrain grid.

Together, these components form a modular framework that links terrain mechanics, uncertainty modeling, and real-time safety assurance. The bounded Bekker formulation captures variability in soil response, the runtime monitor enforces physical limits during operation, and the integration with a vehicle-dynamic planner demonstrates how these ideas can coexist within a practical autonomy stack. The following section presents simulation results that quantify intervention rates, violation probability, and efficiency tradeoffs under different uncertainty bounds and terrain conditions.

\begin{figure}[h!]
    \centering
    \includegraphics[width=1\linewidth]{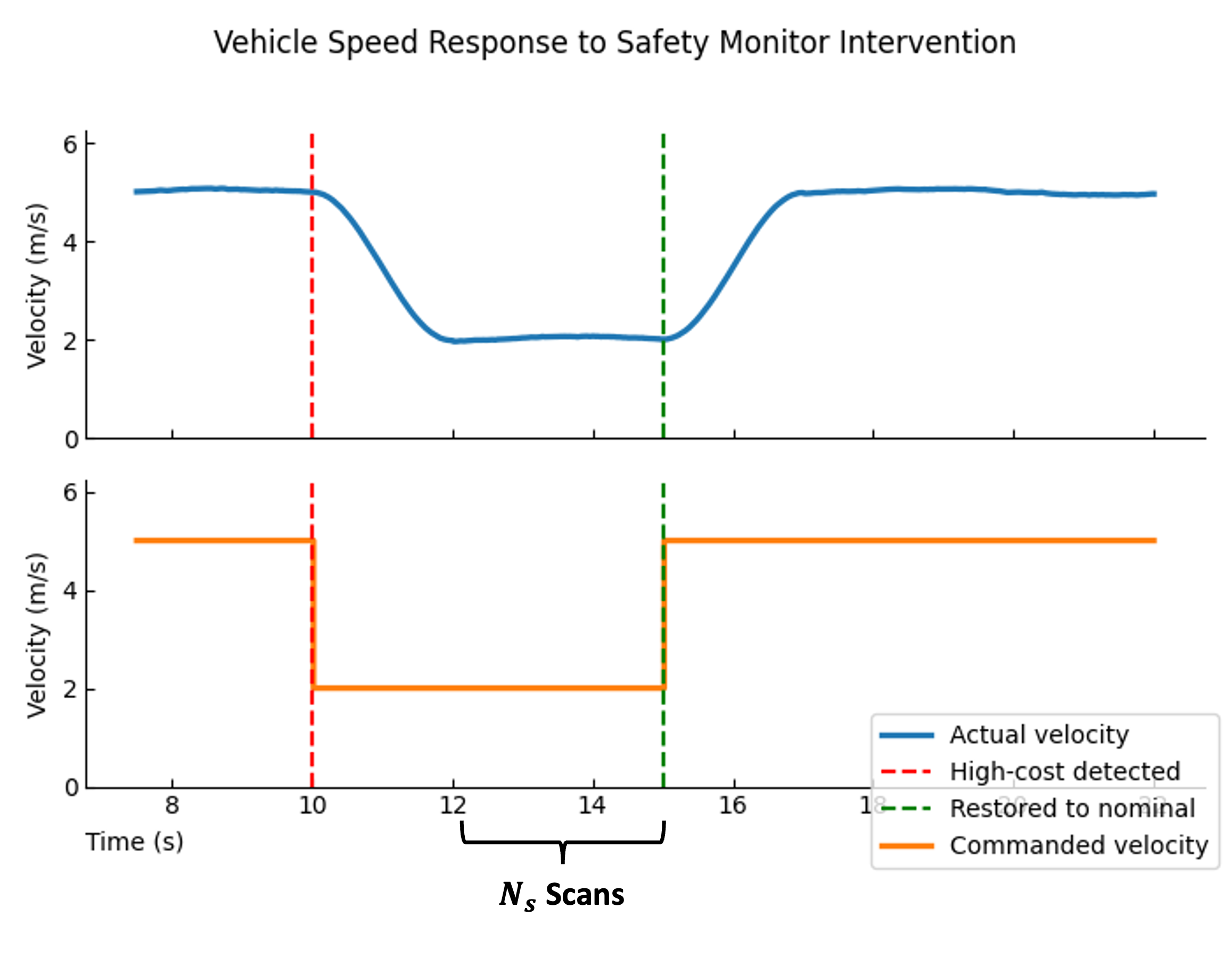}
    \caption{Monitor slowdown behavior: high-cost detected, vehicle slows to $v_{min}=2$\,m/s and restores nominal speed after $N_s=10$ safe updates.}
    \label{fig:velocity_plot}
\end{figure}

\begin{figure}[h!]
    \centering
    \includegraphics[width=1\linewidth]{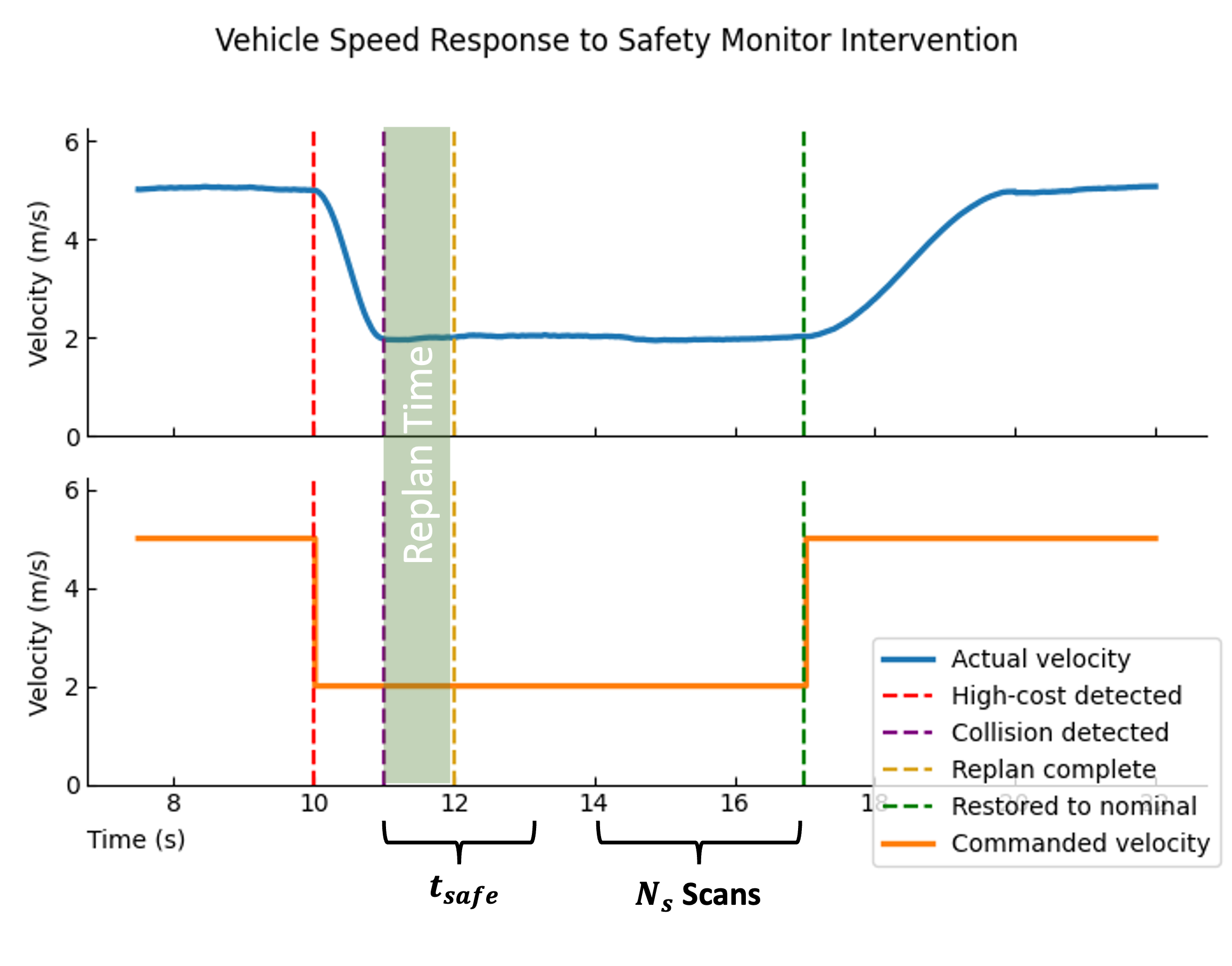}
    \caption{Re-plan within safe window: vehicle maintains low speed until new path is generated, then resumes nominal speed.}
    \label{fig:velocity_plot_replan}
\end{figure}

\begin{figure}[h!]
    \centering
    \includegraphics[width=1\linewidth]{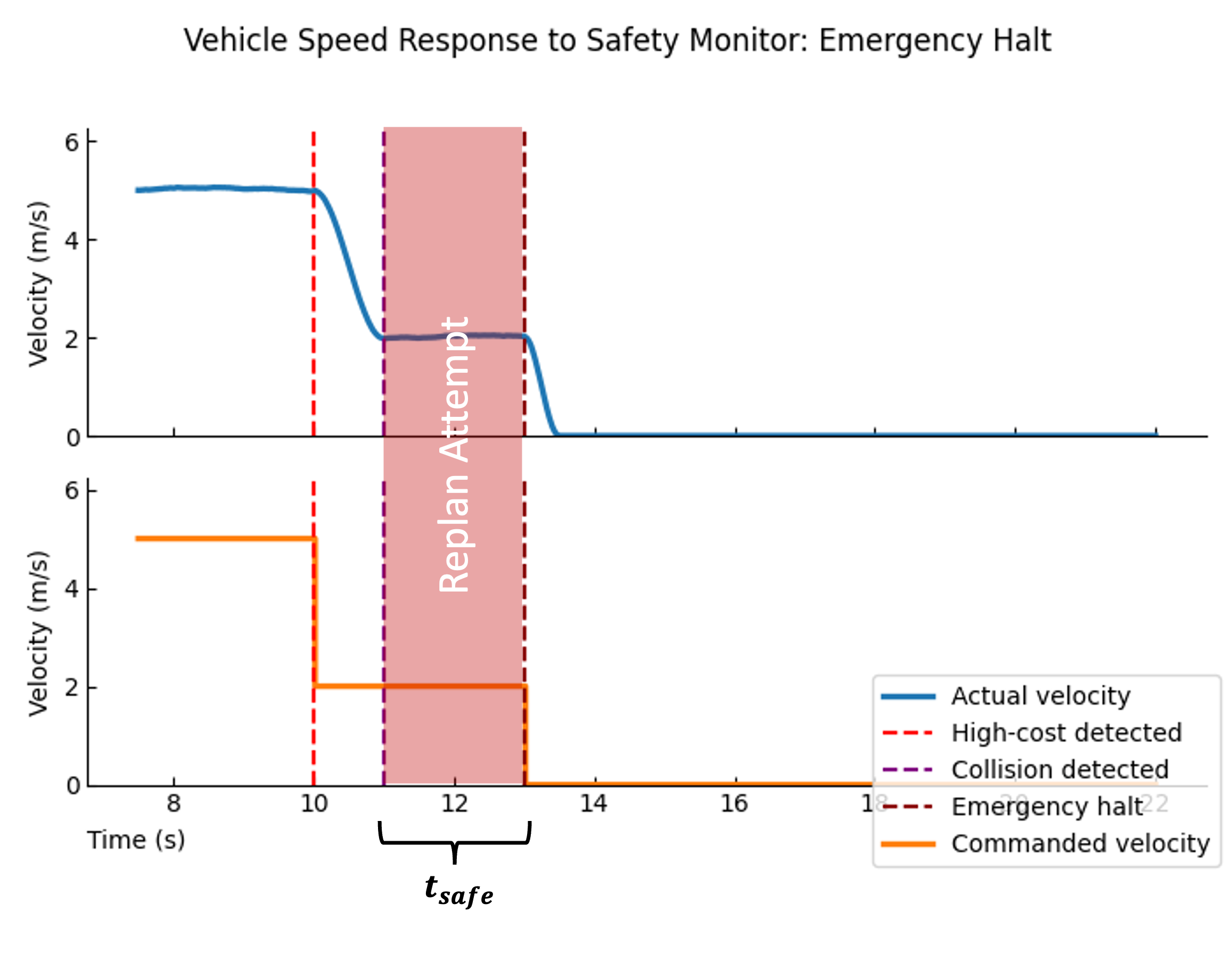}
    \caption{Halt condition: no feasible re-plan before $t_{safe}=2\text{s}$,; vehicle stops at last verified safe cell.}
    \label{fig:velocity_plot_halt}
\end{figure}

\section{Results and Discussion}
The proposed safety monitor was evaluated in a simulated off-road environment $100\text{m}\times100\text{m}$ that combines varying soil types and slopes of the terrain. Each experiment begins with a planner-generated path from (10, 10) to (90, 90). The monitor reacts in real time as new terrain data are revealed through simulated sensing. The results that follow present the terrain model, the behavior of the planner, and the monitor interventions in several uncertainty margins. Figure~\ref{fig:terrain_cost_maps} summarizes the simulated environment used to evaluate the proposed framework. The elevation layer in Figure~\ref{fig:terrain_cost_maps}A defines the local slope used to compute the rollover cost $C_{roll}$. In simulation, elevation is treated as ground-truth data without uncertainty, allowing slope-based penalties to be evaluated deterministically. The soil-type map in Figure~\ref{fig:terrain_cost_maps}B encodes integer labels that correspond to the table of soil parameters used in the Bekker model, mapping each soil class to its respective values $(k_c,\,k_\phi,\,n)$ as listed in Table~\ref{tab:soil_params}. Using these parameters, the bounded Bekker cost $C_{uc}$ shown in Figure~\ref{fig:terrain_cost_maps}C is calculated with an uncertainty margin of $\delta = 0.2$, inflating cost around weaker and less predictable soils. The combined cost in Figure~\ref{fig:terrain_cost_maps}D merges sinkage and slope effects through 
$C_{total} = C_{uc} + \lambda C_{roll}$ with $\lambda = 2.5$, producing continuous high-cost zones that capture both mechanical and geometric terrain risks for use by the motion planner and the runtime safety monitor.

Figure~\ref{fig:trigger_cost_maps} illustrates the entire intervention sequence. As the vehicle moves, new terrain is detected and integrated into the local cost map. The High-cost regions detected within the sensing window are padded with a safety distance $d_{pad}$ calculated from Equation \ref{eq:padding}, ensuring sufficient space for the vehicle to turn or stop before entering the boundary. The available response time before reaching that boundary, $t_{safe}$, is determined using Equation \ref{eq:t_safe}. Figure~\ref{fig:trigger_cost_maps}A shows the true cost revealed by sensing, while Figure~\ref{fig:trigger_cost_maps}B shows the detection of a high-cost region. In Figure~\ref{fig:trigger_cost_maps}C, the region is expanded by the calculated padding distance, forming a safety buffer. Finally, Figure~\ref{fig:trigger_cost_maps}D illustrates a successful re-plan executed within the computed safe-time window. If no feasible detour is found before \(t_{safe}\), the vehicle halts at the last verified safe cell. This mechanism ensures that motion remains within the bounded sinkage and rollover limits defined by the monitor.

\begin{figure}[h!]
    \centering
    \includegraphics[width=0.825\linewidth]{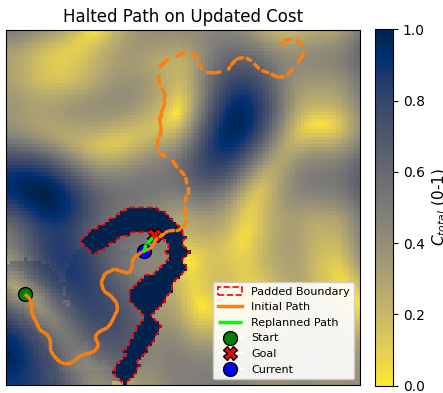}
    \caption{Spatial configuration of a representative halt event. The vehicle stops at the last verified safe waypoint before entering the padded high-cost region.}
    \label{fig:halted_path}
\end{figure}

\begin{table*}[h!]
    \centering
    \small
       \renewcommand{\arraystretch}{1.15}
    \caption{Averaged results over 10 randomized maps (100\,m$\times$100\,m). 
    Each intervention enforces $v_{min}=2$\,m/s. 
    Columns list mean path distance, completion time, normalized cumulative cost, number of high-cost violations, total interventions (speed reductions or re-plans), and number of re-plans. 
    Larger $\delta$ values represent more conservative uncertainty bounds in the Bekker cost model.}
    \label{tab:results-summary}
    \begin{tabular}{>{\centering\arraybackslash}p{0.121\linewidth}>{\centering\arraybackslash}p{0.121\linewidth}>{\centering\arraybackslash}p{0.121\linewidth}>{\centering\arraybackslash}p{0.121\linewidth}>{\centering\arraybackslash}p{0.121\linewidth}>{\centering\arraybackslash}p{0.121\linewidth}>{\centering\arraybackslash}p{0.121\linewidth}}
    \hline \rowcolor{lightgray}
    \textbf{Case} & \textbf{Distance} [m] & \textbf{Time} [s] & \textbf{Cost} & \textbf{Violations} & \textbf{Interventions} & \textbf{Re-plans} \\
    \hline
    Base & 125$\pm$3 & 25.0$\pm$1.2 & 0.69$\pm$0.02 & 12 & 0 & 0 \\
    $\delta$=0.05 & 130$\pm$4 & 27.6$\pm$1.5 & 0.65$\pm$0.02 & 5 & 2 & 1 \\
    $\delta$=0.10 & 135$\pm$5 & 30.8$\pm$1.8 & 0.62$\pm$0.03 & 3 & 4 & 2 \\
    $\delta$=0.15 & 142$\pm$5 & 34.9$\pm$2.0 & 0.59$\pm$0.03 & 2 & 6 & 3 \\
    $\delta$=0.20 & 148$\pm$6 & 39.5$\pm$2.5 & 0.55$\pm$0.03 & 1 & 9 & 4 \\
    $\delta$=0.25 & 156$\pm$6 & 45.8$\pm$2.7 & 0.52$\pm$0.04 & 0 & 12 & 6 \\
    $\delta$=0.30 & 165$\pm$7 & 52.3$\pm$3.0 & 0.50$\pm$0.04 & 0 & 15 & 8 \\
    \bottomrule
    \end{tabular}
\end{table*}

When a high-cost zone enters the sensing range, the vehicle immediately slows to the minimum safe velocity $v_{min}=2 \space \text{m/s}$, regardless of whether the planned path crosses the region. This preemptive slowdown allows for additional sensing updates before reaching the risk boundary. In Figure~\ref{fig:velocity_plot}, the environment remains safe in consecutive cycles. The monitor restores nominal speed after $N_s=10$ safe updates are accumulated. In Figure~\ref{fig:velocity_plot_replan}, a re-plan is triggered within the calculated safe-time window, producing a new path that avoids the padded boundary. This is done while maintaining a safe speed. In Figure~\ref{fig:velocity_plot_halt}, no feasible detour is found before the timeout indicated by $t_{safe}=2\text{s}$, causing the vehicle to halt at the last verified safe cell. This sequence confirms that the monitor can safely stop the vehicle before entering terrain that violates the sinkage and rollover limits. This behavior is further illustrated in Fig.~\ref{fig:halted_path}, which shows the spatial configuration during a representative halt event. In this case, the high-cost region and the padded boundary were deliberately configured to demonstrate stopping logic, as no stops were triggered naturally during the randomized trials summarized in Table~\ref{tab:results-summary}. The figure highlights how the vehicle terminates motion just outside the unsafe zone, consistent with the timeout constraint $t_{safe}$ derived from Eq.~\ref{eq:t_safe}, confirming proactive avoidance before any violation occurs.

Quantitative results in ten generations of randomly selected terrain are summarized in Table~\ref{tab:results-summary}. As the uncertainty margin $\delta$ increases, the monitor enforces more conservative behavior, leading to longer paths and completion times. The mean traversal cost per meter decreases from 0.69 in the baseline VD-RRT* \cite{naik_hybrid_2025} to 0.50 at $\delta=0.30$, indicating that the monitor guides the vehicle to firmer and lower-risk terrain. High-cost violations fall from twelve without monitoring to zero beyond $\delta=0.25$, confirming that bounded Bekker costs maintain terrain safety limits. Interventions and re-plans increase with higher $\delta$ values, showing a natural tradeoff between caution and efficiency. These results indicate that $\delta$ directly tunes the tradeoff between caution and efficiency. Too small a margin leads to frequent high-cost violations, while excessive margins cause unnecessary slowdowns and detours. In all trials, $\delta \approx 0.20$ provided the best balance between intervention frequency and traversal efficiency, representing a practical operating point for the runtime monitor.

Overall, the results show that adding bounded Bekker costs to the runtime safety monitor improves terrain reliability in a clear and measurable way. The monitor keeps motion within safe limits while letting the planner pursue efficient routes, stepping in only when the risk of sinkage or rollover grows too high. Because interventions are localized and physically interpretable, they complement rather than compete with the planner’s objectives. The tradeoff between caution and speed emerges naturally in the form of longer paths, slower completions, but far fewer safety violations. In practice, this balance reflects how uncertainty‐aware design can make off‐road autonomy both safer and more predictable. Future work will focus on validating these behaviors on hardware and refining the uncertainty bounds based on live soil estimation confidence.

\section{Key Implications for Vehicle Design}
Incorporating uncertainty-bounded Bekker models into off-road vehicle design changes how mobility, safety, and autonomy are addressed at the system level. The safety monitor developed in this work links physics-based terramechanics with planning decisions, allowing uncertainty in soil parameters to be treated as a controllable design variable rather than an unpredictable external factor. This approach establishes a direct relationship between ground mechanics and vehicle architecture, allowing design choices that balance performance and reliability in unstructured environments. The framework supports a design philosophy in which safety and mobility margins are defined not as static limits but as dynamic envelopes informed by real-time terrain confidence. A primary implication of this work lies in the design of vehicle powertrain and mobility systems. By quantifying traction uncertainty through Bekker-based cost models, engineers can evaluate how torque distribution, wheel slip, and track engagement respond to variations in soil stiffness and cohesion. These insights can inform the selection of gear ratios, clutch engagement thresholds, and traction control algorithms that maintain mobility under partial traction loss. The uncertainty bounds used on the safety monitor also help determine how much torque reserve or propulsion redundancy is required to maintain performance in variable terrain. This approach aligns the design of mechanical drivetrains with the statistical characteristics of terrain response, improving both energy efficiency and operational predictability.

\begin{table*}[h!]
\centering
\caption{Summary of vehicle design implications derived from the uncertainty-bounded Bekker-based safety monitor.}
\label{tab:design_implications_bekker}
\renewcommand{\arraystretch}{1.5}
\small
\begin{tabular}{
>{\raggedright\arraybackslash}p{0.175\linewidth}
>{\raggedright\arraybackslash}p{0.4\linewidth}
>{\raggedright\arraybackslash}p{0.375\linewidth}}
\hline
\rowcolor{lightgray}
\textbf{Design Domain} & \textbf{Key Implications and Insights} & \textbf{Practical Applications} \\
\hline
Powertrain and traction design & Quantified uncertainty in soil stiffness and cohesion informs torque distribution and traction limits. & Adaptive traction control, torque reserve sizing, drivetrain tuning for variable soils. \\

Chassis and stability systems & Rollover and sinkage margins guide suspension stiffness, center-of-gravity placement, and mass distribution. & Stability envelope design, adjustable suspension calibration, vehicle posture control. \\

Sensor layout and perception & Sensor positioning affects accuracy of terrain parameter estimation and cost confidence. & Ground-contact load sensors, IMU placement, perception-sensor co-design. \\

Energy efficiency and control tuning & Terrain-aware torque control reduces power waste under low traction and high uncertainty. & Predictive torque management, clutch engagement optimization, traction energy mapping. \\

Simulation and validation & Bounded uncertainty enables virtual terrain testing and probabilistic certification metrics. & Monte Carlo terrain evaluation, reliability benchmarking, reduced field trials. \\

Safety assurance and system integration & The monitor connects physical soil models to runtime safety checks, bridging design and operation. & Runtime assurance validation, integrated autonomy safety architecture, certification support. \\
\hline
\end{tabular}
\end{table*}

Another major implication concerns vehicle stability and structural design. The evaluation of the probability of sinkage and rollover by the monitor creates measurable feedback on the distribution of the mass, the stiffness of the suspension, and the clearance of the ground. Designers can use these outputs to assess whether center-of-gravity placement and articulation geometry provide sufficient stability across a full range of soil conditions. For tracked vehicles, where weight distribution strongly affects sinkage, the uncertainty-bounded Bekker model enables early sensitivity studies that reduce the risk of immobilization. The same framework supports adaptive suspension systems and real-time control modes that adjust vehicle posture when the safety margin approaches the defined confidence threshold. The integration of sensors and the design of the perception system also benefit from the proposed approach. The accuracy of terrain parameter estimation depends on where ground-contact sensors, load cells, and inertial units are located. The uncertainty-bounded framework can identify which sensor configurations most effectively reduce variance in cost estimation. By aligning perception layout with terramechanics modeling, the vehicle gains improved observability of soil stiffness and contact pressure, enhancing the precision of planning and stability control. These insights inform future sensor placement standards for off-road autonomous vehicles operating in low-visibility or high-variability terrain.

Finally, the proposed safety monitor influences the validation, testing, and certification processes. The use of bounded uncertainty allows simulation-based evaluations to account for soil variability prior to field deployment, reducing the number of physical trials required. Designers can define mobility envelopes, intervention thresholds, and energy tradeoffs within virtual terramechanics environments, improving confidence in the reliability of the system. This shift toward probabilistic validation supports certification frameworks that prioritize measurable safety margins rather than nominal performance. As autonomy and terrain-aware control become more integrated into off-road platforms, the use of uncertainty-bounded Bekker models will provide a common language linking design, testing, and operational safety across mechanical and control domains.

\section{Conclusions and Future Work}
This paper introduced a runtime safety monitor that enforces terrain safety limits using uncertainty-bounded Bekker costs. By combining classical terramechanics with runtime assurance, the framework ties vehicle safety directly to physical soil response rather than abstract risk metrics. The monitor operates as a lightweight supervisory layer that evaluates terrain costs and intervenes only when the sinkage or rollover limits are likely to be exceeded. The simulation results demonstrated that it maintains safe operation in varying terrain conditions while preserving the efficiency of the planner. Increasing the uncertainty margin led to more conservative but stable behavior, confirming that bounded Bekker costs offer a controllable balance between caution and speed. Future work will focus on hardware validation on a ground vehicle platform to assess the framework under real sensor noise and delay. Ideally, Ground-truth soil measurements would be collected to calibrate Bekker parameters and compare predicted versus observed sinkage. Additional efforts can explore adaptive uncertainty margins that respond to estimation confidence, along with integration with other planners. These steps aim to bridge simulation and field deployment, enabling real-time terrain-aware safety assurance for off-road autonomy.

\section*{Acknowledgments}
The authors declare that they have no conflicts of interest with respect to this work. The authors gratefully acknowledge the Army Corps of Engineers Engineering Research and Development Center, Construction Engineering Research Laboratory, for their review and approval of the final manuscript and financial support under contract award number W9132T23C0013. ChatGPT-5o (OpenAI) and Grammarly were used to assist with editorial work that aimed to improve grammar, wording, and readability during and after the writing process. These tools were not used to generate technical content, collect or analyze literature, create figures, or contribute to the core structure or arguments of the article. The authors have carefully reviewed the final manuscript and take full responsibility for all content and conclusions presented in this work.

\bibliographystyle{ieeetr}
\begin{small}
\bibliography{references}
\end{small}

\end{document}